\documentclass[titlepage]{csetr}

\usepackage{graphicx}

\renewcommand{\and}{\hspace{.5cm}}

\trnumber{201607}
\trdate{April 2016}

\title{%
  Big Data Analytics Using Cloud and Crowd
}
\author{%
  Mohammad Allahbakhsh$^{1,2}$ \and %
  Saeed Arbabi$^2$\\ \and %
  Hamid-Reza Motahari-Nezhad$^{1,3}$ \and %
  Boualem Benatallah$^1$\\[2em]
  $^1\, $University of New South Wales, Australia \\%
  \email{\{mallahbakhsh,hamidm,boualem\}@cse.unsw.edu.au}\\
  $^2\, $University of Zabol, Iran\\%
  \email{\{allahbakhsh,sarbabi\}@uoz.ac.ir}\\
  $^3\,$IBM Almaden Research Center, San Jose, CA, USA \\%
  \email{motahari@us.ibm.com}\\[3cm]
}

\date{}
\begin{document}
\maketitle

\begin{abstract}
The increasing application of social and human-enabled systems in people's daily life from one side and from the other side the fast growth of mobile and smart phones technologies have resulted in generating tremendous amount of data, also referred to as big data, and a need for analyzing these data, i.e., big data analytics. Recently a trend has emerged to incorporate human computing power into big data analytics to solve some shortcomings of existing big data analytics such as dealing with semi or unstructured data. Including crowd into big data analytics creates some new challenges such as security, privacy and availability issues.

In this paper study hybrid human-machine big data analytics and propose a framework to study these systems from crowd involvement point of view. We identify some open issues in the area and propose a set of research directions for the future of big data analytics area.

\end{abstract}

\section{Introduction}
Nowadays, the application of information technology is a vital part of our daily life. People around the globe use billions of mobile devices daily and spend more times on using these digital devices than ever. Sharing our opinions in social networks, searching the web, twitting, purchasing online products, participating in online polling and many other digital aspect of our lives leave behind a tremendous digital footprint. Billions of sensors embedded in cars, mobiles and other forms of devices constantly sense, generate and communicate trillions of bytes of information.
This gigantic generated data, which is also referred to as \emph{big data}, is rich of information about the behavior of individuals in an interconnected network. That is why those who are interested in analyzing human behavior from business analysts to social scientists to academic researchers are highly interested in this data~\cite{IntersBD}.

Decision makers tend to extract individual as well as social behavior indicators from this data in order to make better decisions. Using traditional data management models to process and manage big data is nearly impossible due to the huge volume of data, the vast velocity of data arrival and variety of data types~\cite{ibmBigData}.
Therefore, there is a need to develop special techniques which are able to deal with these aspects of big data in order to support the data-driven decision makings. These techniques are also called \emph{big data analytics}.

Big data management approaches are expected to provide a required level of availability, scalability, security and privacy while working with data~\cite{sherifbigdata2011,IntersBD}. Traditionally, automated techniques are used as big data analytics. Sometimes AI techniques are used to extract information from big data~\cite{supervisedrndwlk}. In some other cases heuristic approaches are used to extract social or individual behavioral indicators from a large community~\cite{FIM}. These techniques while perform reasonably well in some aspect such as storing or retrieving data in cloud data management systems, they might not perform well when it comes to data collection, curation, annotation and dissemination. For example, AI techniques are not able to provide results with very high precisions when working with unstructured or incomplete data~\cite{wallmart}. Also, there are cases in which automated techniques are not able to do the job due to the nature of the tasks. For instance, in a database, there might be some missing data items, such as a person's mail address, that there not exist in the datasets at all, hence no automated technique is able to extract such missing piece of information~\cite{elisa2010curation,crowddb,sortsnjoins}.
To overcome this problem many researches have proposed to enlist the human intelligence and wisdom of crowds in combination with the automated techniques~\cite{wallmart,crowddb,humanOCR,crowdsearcher,sortsnjoins,elisa2010curation,crowdscreen}. Crowdsourcing is a distributed computing method in which, under specific circumstances, can provide contributions comparable to experts' contributions in terms of quality level~\cite{crowdwisdombook,allahbakhsh2013quality,CSModel,crowdscreen}. Crowd involvement in data management tasks, while improves quality of outcomes~\cite{crowdER,crowddb,crowdscreen}, raises new challenges.

In this paper, we first study related-work in the area of big data analytics as well as crowdsourcing.
Then we propose a generic framework that simplifies the analysis of existing hybrid human-machine big data analytics.
The result of such an analysis is a set of problems that are yet to be answered.
We propose such set of challenges and propose some directions for future research in the area.

In summary, In Section~\ref{sec:rels}, we study related work in the area of big data analytics and crowdsourcing. In Section~\ref{sec:frm}, we propose our analysis framework. The open issues are studied in Section~\ref{sec:issues}, and we conclude in Section~\ref{sec:concs}.

\section{Related Work}\label{sec:rels}
We organize this section in three different sub-sections. We first study big data analytics. We then study the crowdsourcing basic concepts and finally we use the Wall-Mart case study to articulate the problems that need more investigations.

\subsection{Big Data Analytics}
Many systems such as social networks, sensing systems, etc., produce very large amounts of information. This data is not called big data only because of its size. Four important attributes, also referred to as \emph{4V}, characterize the big data concept:(i)~ data is huge in terms of its \emph{volume}; (ii)~data is produced with a very high \emph{velocity}; (iii)~data comes from a great \emph{variety} of data types; and finally (iv)~data has different levels of \emph{veracity}.
Such a tremendous volume of data is a rich source of information about the behavior of individuals, social relations between individuals, patterns, e.g., purchase patterns, in the behavior of individuals and so on. Hence, extracting these hidden aspects is of a great importance to the business owners and analysts. The process of extracting these information from big data is called big data analytics and are applied using different techniques and methods~\cite{IntersBD,sherifbigdata2011,AnalyticsonBD,bigdatareport,bda2012}.

With the rise of recent web technologies and especially emergence of Web 3.0, recent applications which are working with big data aim to be implemented as distributed, scalable and widely accessible service on the web. Cloud computing paradigm makes applications available as services from anywhere in the world by shifting the infrastructure to the network. The following properties of cloud computing has made it a good candidate for hosting deployments of data-intensive applications:

-It produces virtually unlimited capacity by providing means to consume the amount of IT resources that is actually needed.

-It reduces costs by only paying for what you use (pay-as-you-go).

-It reduces the time that IT systems have to spend on managing and supporting infrastructure.

For example in 2007 New York Times aimed to build a service for users to have access to any New York Times issue since 1851, a service called TimesMachine. The big challenge was serving a bulk of 11 millions of articles in the form of PDF files. To process these 4 TeraBytes of files they decided to use Amazon Elastic Compute Cloud (EC2) and Simple Storage Service (S3). The source data was uploaded to S3 and then a cluster of Hadoop EC2 Amazon Machine Images (AMIs) was started. With parallel running of 100 EC2 AMIs, the task of reading the source data from S3, converting it to PDF and storing it back to S3 was completed within 36 hours~\cite{zaharia2008improving}.

At the beginning, big data analytics started to be done using the existing advanced analytics disciplines such as data mining techniques. Authors in~\cite{supervisedrndwlk} use supervised learning techniques for link prediction in social networks. As another example, Grahne et. al. propose method for mining sets of items that have been frequently purchased together~\cite{FIM}. Since the clique detection techniques in NP-Hard by nature, Grahne and his colleagues propose a heuristic method to partially overcome this problem. Authors in~\cite{mainWWW} use the same itemset mining technique for detecting spam groups in consumer review systems. The first author and his colleagues have also used mining techniques in~\cite{MohammadCompSec,CollusionMohammad} and iterative techniques in~\cite{RTV,TPDS} to identify collusion groups in the log of online rating systems.

Traditional data management techniques may work fine as long as data are structured. However, one of the main characteristics of big data is the wide variety of data types. In most of the cases data are unstructured or incomplete. Hence, existing data mining or management techniques cannot handle big data. To solve this issue big data management systems have been proposed. Current big data management systems in cloud use some well-known application logics can be categorized to MapReduce, SQL-Like and Hybrid~\cite{sherifbigdata2011}:

The most well-known big data management system is Apache Hadoop\footnote{http://hadoop.apache.org/}, a framework for running data intensive applications on clusters of commodity hardware. Hadoop, which has been very successful and widely used in industry and academia, is a an open source java implementation of MapReduce~\cite{mapreduce}. MapReduce is a simple but powerful programming model that is designed to enable programmers to develop scalable data-intensive applications on clusters of commodity PCs~\cite{mapreduce,sherifbigdata2011}. The MapReduce model is inspired by the map and the reduce functions in functional programming languages. A map instruction partitions a computational task into some smaller sub-tasks. These sub-tasks are executed in the system in parallel. A reduce function is used to collect and integrate all results from sub-tasks in order to build up the main task's outcome. More than 80 companies and organizations (e.g. AOL, LinkedIn, Twitter, Adobe, VISA) are using Hadoop for analytic of their large scale data. Some efforts have been done to add SQL-Like flavor on top of MapReduce as many programmers would prefer working with SQL as a high-level declarative language instead of low-level procedural programming using MapReduce. Pig Latin\cite{PigLatin} and Sawzal~\cite{Sawzall} are examples of such tools.  Finally, Some Systems have been designed with the main goal of bringing some familiar relational database concepts (such as. tables and columns) and a subset of SQL to unstructured world of Hadoop while enjoying Hadoop's the extensibility and flexibility. An example of hybrid solutions is HadoopDB project that tries to combine the scalability of MapReduce with the performance and efficiency of parallel databases.

In parallel with recent trend that convinces companies to give up building and managing their own data centers by using computing capacity of cloud providers, many companies are willing to outsource some of their jobs given the low costs of transferring data over the internet and high costs of managing complicated hardware and software building blocks. Therefore Amazon concluded that cloud computing can allow having access to a workforce that is based around the world and is able to do things that computer algorithms are not really good for. Therefore Amazon launched Machanical Turk (MTurk) system as a crowdsourcing internet marketplace where now has over 200K workers in 100 different countries.

\subsection{Crowdsourcing}
Crowdsourcing is the process of enlisting a crowd of people to solve a problem~\cite{cswww,csfirst}. The idea of crowdsourcing was introduced first by Jeff. Howe in 2006~\cite{csfirst}. Since then, an enormous amount of efforts from both academia and industry has been put into this area and so many crowdsourcing platforms and research prototypes (either general or special purpose)have been introduced. Amazon Mechanical Turk\footnote{http://www.mturk.com}(MTurk), Crowdflower\footnote{http://www.crowdflower.com}, Wikipedia\footnote{http://www.wikipedia.org} and Stackoverflow\footnote{http://stackoverflow.com} are examples of well-known crowdsourcing platforms.

To crowdsource a problem, the problem owner, also called the \emph{requester}, prepares a request for crowd's contributions and submits it to a crowdsourcing platform. This request, also referred to as the \emph{crowdsourcing task} or shortly as the \emph{task}, consists of a description of the problem  that is asked to be solved, a set of requirements necessary for task accomplishment, a possible criteria for evaluating quality of crowd contributions and any other information which can help workers produce contributions of higher quality levels. People who are willing to contribute to the task, also called \emph{workers}, select the task, if they are eligible to do so, and provide the requester with their contributions. The contributions are sent to the requester directly or through the crowdsourcing platform. The requester may evaluate the contributions and reward the workers whose contributions have been accepted~\cite{allahbakhsh2013quality,towardCS,CSforEnter}.

Several dimensions characterized a crowdsourcing task, each of which impacting various aspects of the task from outcome quality to execution time or the costs.

\subsubsection{Task Definition.}
Task definition is important in the success of a crowdsourcing process. a poorly designed task can result in receiving low quality contributions, attracting malicious workers or leaving the task unsolved due to unnecessary complications~\cite{AMTOpportunities,growingfield}. Therefore, it is highly recommended to design robust tasks. A robust task is designed so that it is easier to do it rather than to cheat~\cite{allahbakhsh2013quality}. Moreover, a requester should make sure that she has provided the workers with all information required for doing the task to increase the chance of receiving contributions of higher quality levels. The importance of this dimension is because of its direct impact mainly on the outcome quality, task execution time and number of recruited workers.

\subsubsection{Worker Selection.}

Quality of workers who contribute to a task can directly impact the quality of its outcome~\cite{allahbakhsh2013quality,growingfield,Haleh2013recruitment}. Low quality or malicious workers can produce low quality contributions and consequently waste the resources of the requester. Research shows that recruiting suitable workers can lead to receiving high quality contributions~\cite{Haleh2013recruitment,HalehRandomWalk}. A suitable worker is a worker whose profile, history, experiences and expertise highly matches the requirements of a task.
In a crowdsourcing process, workers might be recruited through various methods such as Open-call, Publish/Subscribe~\cite{twitter,cssmart}, Friend-based~\cite{friendsourcing}, profile-based~\cite{Haleh2013recruitment} and team-based~\cite{peoplecloude}.

\subsubsection{Real-time Control and Support.}
During the execution of the task, the requester may manually or automatically control the workflow of the task and manipulate the workflow or the list of the workers who are involved in the task in order to increase the chance of receiving high quality contributions~\cite{turkomatic,crowdweaver}. Moreover, workers may increase their experience while contributing to a task by receiving real-time feedback from other workers or requester. The feedback received in real-time, and before final submission of the worker's contribution, can assist her with pre-assessing her contribution and change it so that satisfies the task requirements~\cite{shepherding}. Real-time workflow control and giving feedback can directly impact the outcome quality, the execution time and also the cost of the task, so they should be taken into account when studying crowdsourcing processes.

\subsubsection{Quality Assessment.}
Assessing the quality of contributions received from the crowd is another important aspect of a crowdsourcing process. Quality in crowdsourcing is always under question. The reason is that workers in crowdsourcing systems have different levels of expertise and experiences; they contribute with different incentives and motivations; and even they might be included in collaborative unfair activities~\cite{allahbakhsh2013quality,allahbakhsh2012reputation,turf}.Several approaches are proposed to assess quality of workers' contributions such as expert review, Input agreement, output agreement, majority consensus and ground truth~\cite{allahbakhsh2013quality}.

\subsubsection{Compensation Policy.}
Rewarding the workers whose contributions have been accepted or punishing malicious or low quality workers can directly impact their chance, eligibility and motivation to contribute to  the future tasks. Rewards can be monetary (extrinsic) or non-monetary (intrinsic). Research shows that the impact of intrinsic rewards, e.g., altruism or recognition in the community, on the quality of the workers' contributions is more than the monetary rewards~\cite{intrinsic}. Choosing an adequate compensation policy can greatly impact the number of contributing workers as well as the quality of their contributions. Hence, compensation policy is an important aspect of a crowdsourcing process.

\subsubsection{Aggregation Technique.}
A single crowdsourcing task might be assigned to several workers. The final outcome of such a task can be one or few f the individual contributions received from workers or an aggregation of all of them~\cite{towardCS,CSforEnter}. Voting is an example of the tasks that crowd contributions are aggregated to build up the final task outcome. In contrast, in competition tasks only one or few workers' contributions are accepted and rewarded. Each of the individual contributions has its own characteristics such as quality level, worker's reputation and expertise and so many other attributes. Therefore, combining or even comparing these contributions is a challenging tasks and choosing a wrong aggregation method can directly impact the quality of the task outcome.

\section{Taxonomy of Hybrid Human-Machine Big Data Analytics}\label{sec:frm}

 In this section, we first propose an overview of the concept of combining crowd and big data analytics. We then propose a framework in order to simplify understanding and studying hybrid human-machine big data approaches.

\subsection{Overview}
As mentioned earlier, one of the main characteristics of big data is the wide variety of data types. Data might be from different types; they might be semi-structured, unstructured or structured but incomplete. Traditional or advanced analytics cannot handle such a variable tremendously large data. So, they should be handled using the big data analytics. As we studied in the previous section, generally, the big data analytics rely on the deterministic or learning techniques which use the computing power of machines to process big data and extract necessary information, patterns, etc. Due to huge size and high level of complexity of big data, the techniques used for data analysis usually leverage heuristic or learning techniques, and hence, they are not able to guarantee their performance. In some cases the requester might need a predefined specific level of precision and accuracy for the results, but the existing techniques cannot provide that level of precision.

In these cases, to generate results having the specified quality level a crowd of people are employed to complement the performance of the machines. To simplify understanding these concepts we study CrowdER~\cite{crowdER} and WallMart crowdsourcing project~\cite{wallmart} as two related work. We study these systems later in Section~\ref{sec:dim} to study how they deal with the inclusion of crowd into cloud.

CrowdER is a hybrid crowd-machine approach for entity resolution. CrowdER first employs machines to perform an initial analysis on the data and find most likely solutions and then employs people to refine the results generated by machines.
Wall-Mart product classification project is another example of hybrid-crowd-machine approaches proposed for big data analytics~\cite{wallmart}. In this project, a huge volume of data is constantly being received from various retailers from all across the country. The sent data is structured but in most of the cases it is incomplete. So, Wall-Mart cannot use only machines for the purpose of entity matching and resolution. The results, as tested, do not have the required level of accuracy. To overcome this problem, Wall-Mart selects some individuals with adequate expertise to refine the results produced by machine.

\subsection{Dimensions}
\label{sec:dim}

\begin{figure*}[!t]
\centering
\includegraphics[scale=0.45]{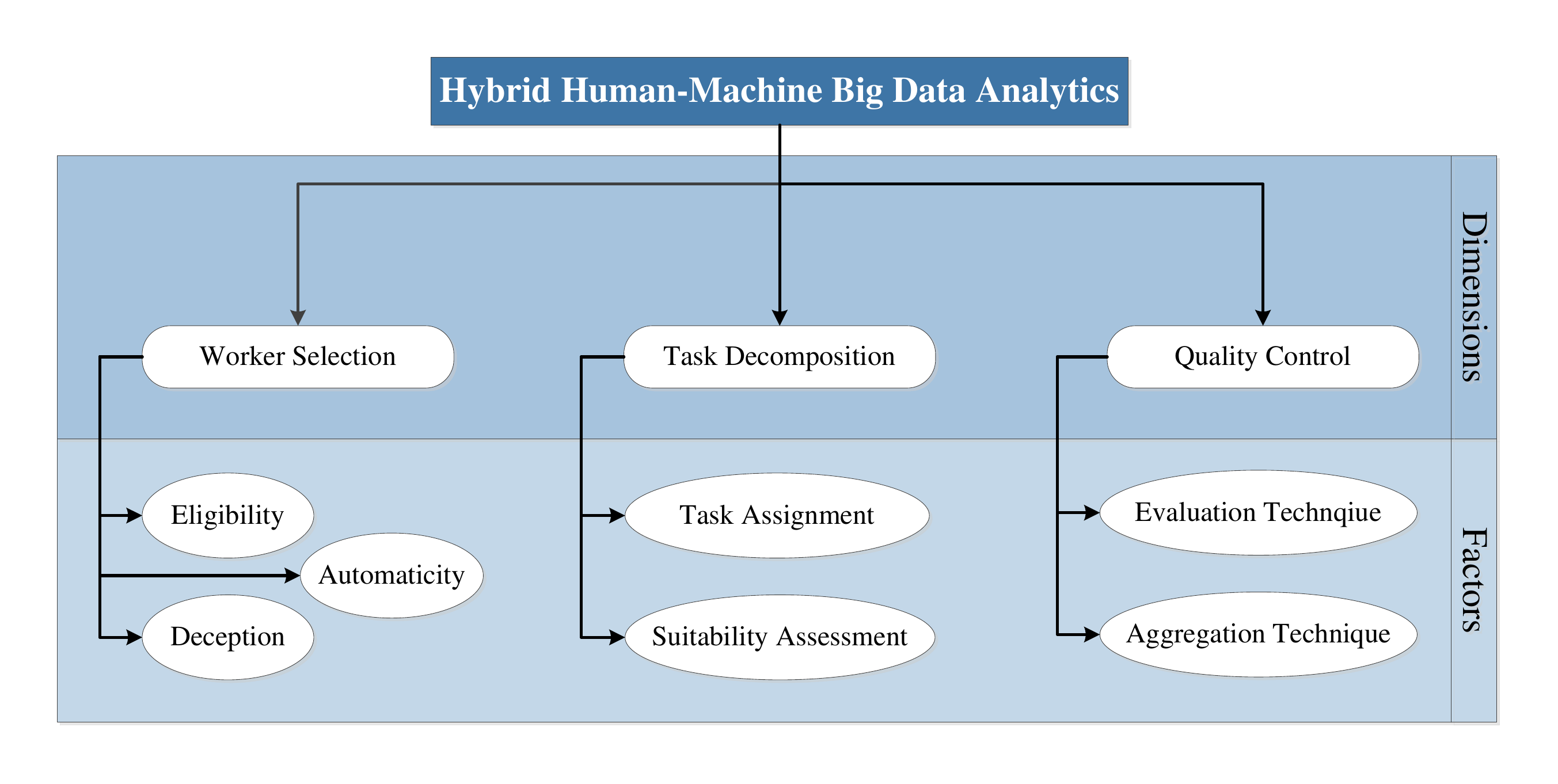}
\caption[]{Taxonomy of Hybrid Human-Machine Big Data Analytics}
\label{fig:tax}
\end{figure*}

Big data analytics can be characterized using various dimensions. Some of these dimensions are related to the nature of big data and how the machine-based approaches deal with it. Availability, multi-tenancy, scalability, etc. are examples of these dimensions. These dimensions are well studied in several related work`\cite{sherifbigdata2011,bda2012,IntersBD}. In this work, we do not study these dimensions and identify new dimensions that emerge when a crowd of people in added to big data analytics. From this view point, we identify three important dimensions: \emph{worker selection}, \emph{task decomposition} and \emph{quality control}. In the followings, we describe these dimensions and study CrowdER and Wall-Mart project as two existing case studies from point of view of these dimensions. Figure~\ref{fig:tax} depicts a representation of the identified dimensions and important factors that should be taken into account when studying each dimension.

\subsubsection{Worker Selection. }
When somebody tries to empower big data analytics with a crowd of workers, there are several points that she should have in mind.  The first point is how to assess eligibility of workers. Doing some big data analyses may need specific skills and expertise. In such cases only workers having suitable profiles containing those expertise should be selected. Defining good worker selection policies can greatly impact quality of analytics. In Wall-Mart project, since workers are selected manually from a private crowd, their profiles are checked against the requirements and only suitable workers are chosen. In CrowdER no worker selection policies are put in place and everybody can contribute to the data analysis process.

The second point in worker selection is that, workers should be selected so that the chance of deception is decreased. Human-enhanced systems are constantly subject to misbehavior and deceptive activities. The worker selection should be applied so that both individual and collaborative deceptive activities have a very low chance to happen. Selecting untrustworthy workers may result in producing dubious low quality results, so it is crucial to select trustworthy workers. CrowdER does not apply any trust assessment technique while selecting workers, but the Wall-Mart project people are selected from a private crowd, so they generally have a minimum level of trustworthiness.

The third parameter is automaticity, i.e., how to approach and select workers. In some systems, workers are selected manually. In this case the system or task owner looks for people with suitable profiles and asks them to contribute to her task. Wall-Mart project follows a manual worker selection method. On the other hand, in automatic methods, system on behalf of the task owner and, based on the task requirements and people profile, selects and recruits a group of suitable workers. CrowdER recruits its required workers automatically from people at the Amazon Mechanical Turk.

\subsubsection{Task Decomposition. }
By task decomposition, we mean the process of breaking down a  complex analytical task into some smaller sub-tasks that are easier to accomplish mainly in parallel. These sub-tasks should be divided among the machines and humans involved in the system. The first factor is how to decide on ``who should do what?". More precisely, what is the basis on which the task owner or the system decides to assign a sub-task to a human or a machine to do it. Task assignment should be done based on several factors. The first factor is the type of the task. Computers are good in performing computational tasks, but there are tasks that require some levels of intelligence to be done. There are several tasks types that fit in this category such as image tagging, photo description, audio transcription and so on. These tasks are better to be assigned to a human. The type of the task is an important factor to decide to assign the task to a machine or a human. In both CrowdER and Wall-Mart projects, tasks are first assigned to machines to do. Then, the results obtained from machines are given to people to refine the results so that they meet the precision requirements.

After a task owner decides to assign a task to humans, the next challenges emerges: ``who is suitable to do the task?" Several methods are proposed for eligibility assessment to answer this question~\cite{Haleh2013recruitment,allahbakhsh2012reputation,HalehRandomWalk,allahbakhsh2013quality}. Generally speaking, the humans whose profiles have a higher level of match with the task requirements are more suitable workers to participate in the task. Wall-Mart selects people manually and they check the suitability of workers when they select them to participate, but in CrowdER no suitability assessment approaches are used and everybody can contribute to the tasks.

\subsubsection{Quality Control. }
By quality control, we mean all the activities performed to make sure that the outcome of a hybrid human-machine big data task has an accepted level of quality. Since, the task is performed by both machines and human and then the individual contributions are aggregated to build up the final task outcome, quality of the final task's outcome depends on two parameters: quality of the individual contributions and the aggregation algorithm.

Quality of each single contribution is an important factor that directly impacts the quality of the final task outcome. Quality of humans' contributions depends on many parameters such as quality of the contributor, task requirements, the process of contribution generation (e.g., time of the process, cost of the process, etc.) and so on~\cite{allahbakhsh2013quality}. Quality of machines' contributions mainly depends on the algorithm that is used to generate the contributions. Many quality control approaches are proposed for assessing quality of contributions in human-involved tasks such as expert review, ground truth, voting, input agreement, output agreement, etc.~\cite{allahbakhsh2012reputation}. Wall-Mart project uses the expert review method for assessing quality of single contributions received from machines. But they do not control the quality of contributions received from human, because of the way they have recruited them. In CrowdER, the ground truth method is used to assess quality of machines' contributions. Then, the contributions passing a minimum threshold are sent to crowd for final assessment. To increase the chance of receiving quality results from humans, the CrowdER, encourages them to provide a reason why they think the entities match. This is a quality assurance (design time) technique called robust task design~\cite{allahbakhsh2013quality}.

The aggregation algorithm is another important parameter impacting quality of the final outcome. For the simplest form in which all contributions are generated by only humans or machines, the aggregation algorithms take into account many parameters such as the quality of individual contributions, quality of the contributor, statistical distribution of contributions, etc.~\cite{allahbakhsh2012reputation,allahbakhsh2013quality,myWWW,RTV,TPDS}. In the context of hybrid human-machine systems the situation is more complex. That is because, in addition to all those parameters, a priority between machine and human should be considered as well. Assume that in a special case, a disagreement occurs between the human response and machine's result. Which one should be selected as the correct result?~\cite{cswww} In almost all existing hybrid systems, the priority is given to human's response whenever a contradiction happens. This is the case in CrowdER and Wall-Mart project as well. In both cases the initial results are generated by machines and refined and curated by humans.

\section{Open Issues}\label{sec:issues}

The hybrid big data analytics is a young but fast growing research area. Billions of people all around the world are equipped with mobile devices all capable of creating gigantic amounts of unstructured data forces the big data analytics area to leverage human computational power towards analyzing big data. The involvement of humans in the big data analytics, while resolves some challenges, creates new challenges that need more attention and investigations. In this section we bring some of these challenges and discuss how they probably can be solved in future.
\subsection{Task Decomposition}
Task decomposition as described in Section~\ref{sec:frm} is an important dimension in hybrid human-machine big data analytics. Task decomposition has several challenging aspects.
The first aspect is how to decompose the task. Should the task be decomposed automatically or humans should be involved in? Automated task decomposition is faster, but probably not enough accurate. On the other hand, involving public crowd with different levels of knowledge and experience as well as various and sometimes dubious incentives and motivations is a risky job that may waste time and money of the task owners. Employing expert crowd to do the decomposition is another solution which incurs higher costs, may cause more delays, etc. At the moment, the existing hybrid systems employ the manual (expert-based) approaches for decomposing the tasks.

The next challenge is assigning decomposed sub-tasks to humans and machines. Deciding on ``who should do what" is seriously challenging. The first challenge is to decide which part of the work should be assigned to machine and which parts should be given to crowd to be done. The next challenging aspect here is crowd selection. Assigning sub-tasks to crowd can easily create security and privacy problems. There are cases in which assigning whole the problem to one worker can cause serious privacy and security problems~\cite{humanOCR}. Even these problems may occur if a large number of tasks of a requester goes to one worker, or a small group of collaborating workers, the security and privacy of the requester easily breach~\cite{HalehRandomWalk,Haleh2013recruitment}. Finding a privacy aware worker recruitment approach which is also security aware as too and in the same times has a reasonable performance can be a potential direction of research in the future of hybrid human-machine big data analytics.

Furthermore, considering the size of the big data, the high velocity of data arrival and unstructured nature of big data makes the problem more complicated. Let's use an example to explain the challenge. Assume that in a big data management system people are employed to extract entities from a given text. As the size of the problem is too big, a very large number of workers are required to do the job. Assume that the simple task is extracting entities from a given text. There are several approaches to allocate tasks to users~\cite{Haleh2013recruitment,HalehRandomWalk}. Letting every worker to see all tasks simply can overwhelm workers with stacks of tasks and it makes it impossible for them to find the jobs that suits their expertise. So, it is necessary to have a mechanism for assigning tasks to workers. Assigning tasks according to the workers' profiles is not also possible, because the types of data may change suddenly and system should change the worker selection criteria on the fly to match the new requirements. Also data is unstructured, hence, matching the data with the profile of the workers in real-time is another challenge that needs investigations. Moreover, the unstructured nature of big data makes it challenging to use publish/subscribe model for worker selection. In publish/subscribe approach, workers apply to receive updates or tasks from a specific category but because big data is usually unstructured it is hard to fit arrived data items in particular predefined categories. So it is not possible to assign tasks to workers based on this approach.
Finding a task assignment approach is an interesting and challenging aspect of human-enhanced big data analytics.

\subsection{Aggregation Techniques}
The final outcome of a hybrid big data computation task is an aggregate of all contributions received from both machines and humans. Selecting a inadequate aggregation algorithm can easily lead to producing low quality results. The size of the big data is very large; the data is rapidly changing and unstructured. All these characteristics emphasize that the traditional aggregation algorithms are unable to handle this problem. Simple algorithms even, the ones that rely on automated methods and do not involve humans in their computations such as~\cite{RTV} and~\cite{AleksRep}, cannot show an accepted performance in the area.
On the other hand, when it comes to crowd contributions, quality is always a serious question. People have the potential to exploit the system in order to gain unfair benefits and they do cleverly so that they can easily trap any aggregation algorithm~\cite{RepSurvey,reptrap,IEEE2102Survey}.  Selecting a suitable aggregation algorithm is a serious challenge even in traditional big data algorithms.

The combination of human and machine raises even more challenges. In some cases, such as computation tasks, the credibility of the results generated by machines is far more than human results, while in HITs the credibility of crowd's contributions is much higher than machine's.
Also, there might be cases in which there are contradictions between the machine's outcome and a worker's contribution. Which one should have the higher priority in this case? Machine or human? Answering these questions needs deeper investigations and research supported with excessive experimentations.

\subsection{Availability}
Availability of service is one of the main attributes of cloud and big data services~\cite{sherifbigdata2011}. According to the agreements between the service provider and the customer, the system is responsible for providing on-demand resources for the requester. The existing techniques have solved the availability problem reasonably well~\cite{cloudTax,sherifbigdata2011}.
On the other hand, the crowd workers usually are not full-time workers and do not have obligations to be on-time, on-call or even available in a specific period of time~\cite{CSsystems}. Some systems have solved this problem by paying workers to be on-call~\cite{RealtimeCS}. Considering the size of the problems in big data, paying on-call workers drastically increases the costs and is not feasible to apply to this area.

Moreover, in addition to the size of the data, big data is unstructured and also may come from disparate data sources~\cite{BigdataLake}. Therefore, analyzing such a diverse data in terms of type requires having a large crowd of on-call workers with a very broad range of expertise. Providing and keeping on-call such a large crowd having diverse set of expertise is a challenging task which in most of cases is not feasible, or at least imposes high costs to the system.

Workers may come and get engaged in a task, then suddenly leave it without any notice or clear reason. This crowd characteristic raises another challenge.  Assume that a worker has started a task, e.g., tagging data in an online stream, and then suddenly leaves the system due to a communication problem or intentionally. What would be the best scenario to solve this issue? How is the best worker whom should be replaced with the left worker? Should the task be restarted from beginning or it should resume executing from a specific milestone? How these milestones should be defined? And many other problems which all emerge because of leaving and joining workers in a human-enhanced big data management system.

All these crowd characteristics raise a great concern on the availability of service in hybrid big data systems. Therefore, availability is probably another promising and interesting future research direction in the area of hybrid big data.

\section{Conclusion and Future Directions}\label{sec:concs}
The collaborative and social nature of Web 2.0has created an opportunity for emergence of a wide variety of human-centric systems, such as social networks, blogs, online markets, etc., in which people spend a large portion of their daily life. The fast growth of smart phones has made these applications more available than ever. This has led to the era of big data where users deal with data of a tremendous volume, wide variety of types, a rapid velocity and variable veracity. To understand this gigantic era, users need to rely on big data analytics to extract useful information from such a huge pool of data.
Recently, a trend has emerged towards analyzing big data using human intelligence and wisdom of the crowds. This trend, creates great opportunities for big data analytics to create more accurate analytics and results, however they raise many challenges that need more research and investigations.

In this paper, we have studied the area of hybrid human-machine big data analytics. We first studied the crowdsourcing and big data areas and then proposed a framework to study the existing hybrid big data analytics using this framework. We then introduced some significant challenges we believe to point to new research directions in the area of hybrid big data analytics.

We believe that in the near future the trend towards hybrid systems which contain human as a computing power will dramatically increase and a great amount of efforts from both industry and academia will be put in studying hybrid human-enabled systems. As many researches show the emergences of these human-centric systems has created a great challenge on the security and privacy of future big data services. Also, the existence dominance of casual workers in these hybrid systems makes it harder than ever to guarantee availability and quality of services. Social networks, global crowdsourcing systems and other types of mass collaborative systems on the web are pools rich of work forces that can easily participate in big data analytics. At the moment these systems are almost isolated and it is very hard to define a task on all of these pools of human resources. We believe that in the future it will be a need for a generic people management system on the web which is capable of handling profiles, records and histories of people all around the web and acts as a broker for all systems which request for human involvement. Such a people management system can simplifies handling security and privacy issues as well as finding available crowd workers for tasks which need high levels of availability.

\bibliographystyle{plain}

\begin{thebibliography}{10}

\bibitem{allahbakhsh2013quality}
Mohammad Allahbakhsh, Boualem Benatallah, Aleksandar Ignjatovic, Hamid~Reza
  Motahari-Nezhad, Elisa Bertino, and Schahram Dustdar.
\newblock Quality control in crowdsourcing systems: Issues and directions.
\newblock {\em Internet Computing, IEEE}, 17(2):76--81, 2013.

\bibitem{TPDS}
Mohammad Allahbakhsh and Alex Ignatovic.
\newblock An iterative method for calculating robust rating scores.
\newblock {\em IEEE Transactions on Parallel and Distributed Systems}, page~1,
  2014.

\bibitem{RTV}
Mohammad Allahbakhsh and Aleksandar Ignjatovic.
\newblock Rating through voting: An iterative method for robust rating.
\newblock {\em CoRR}, abs/1211.0390, 2012.

\bibitem{allahbakhsh2012reputation}
Mohammad Allahbakhsh, Aleksandar Ignjatovic, Boualem Benatallah,
  Seyed-Mehdi-Reza Beheshti, Elisa Bertino, and Norman Foo.
\newblock Reputation management in crowdsourcing systems.
\newblock In {\em Collaborative Computing: Networking, Applications and
  Worksharing (CollaborateCom), 2012 8th International Conference on}, pages
  664--671. IEEE, 2012.

\bibitem{CollusionMohammad}
Mohammad Allahbakhsh, Aleksandar Ignjatovic, Boualem Benatallah,
  Seyed-Mehdi-Reza Beheshti, Elisa Bertino, and Norman Foo.
\newblock Collusion detection in online rating systems.
\newblock In {\em Proceedings of the 15th Asia Pacific Web Conference (APWeb
  2013)}, pages 196--207, 2013.

\bibitem{MohammadCompSec}
Mohammad Allahbakhsh, Aleksandar Ignjatovic, Boualem Benatallah,
  Seyed-Mehdi-Reza Beheshti, Norman Foo, and Elisa Bertino.
\newblock Representation and querying of unfair evaluations in social rating
  systems.
\newblock {\em Computers \& Security}, 2013.

\bibitem{myWWW}
Mohammad Allahbakhsh, Aleksandar Ignjatovic, HamidReza Motahari-Nezhad, and
  Boualem Benatallah.
\newblock Robust evaluation of products and reviewers in social rating systems.
\newblock {\em World Wide Web}, pages 1--37, 2013.

\bibitem{Haleh2013recruitment}
H.~Amintoosi and S.S. Kanhere.
\newblock A trust-based recruitment framework for multi-hop social
  participatory sensing.
\newblock In {\em Distributed Computing in Sensor Systems (DCOSS), 2013 IEEE
  International Conference on}, pages 266--273, May 2013.

\bibitem{HalehRandomWalk}
Haleh Amintoosi, Salil~S. Kanhere, and Mohammad Allahbakhsh.
\newblock Trust-based privacy-aware participant selection in social
  participatory sensing.
\newblock {\em Journal of Information Security and Applications}, (0):--, 2014.

\bibitem{supervisedrndwlk}
Lars Backstrom and Jure Leskovec.
\newblock Supervised random walks: Predicting and recommending links in social
  networks.
\newblock In {\em Proceedings of the Fourth ACM International Conference on Web
  Search and Data Mining}, WSDM '11, pages 635--644, New York, NY, USA, 2011.
  ACM.

\bibitem{CSsystems}
Michael~S Bernstein.
\newblock Crowd-powered systems.
\newblock {\em KI-K{\"u}nstliche Intelligenz}, 27(1):69--73, 2013.

\bibitem{RealtimeCS}
Michael~S. Bernstein, Joel Brandt, Robert~C. Miller, and David~R. Karger.
\newblock Crowds in two seconds: Enabling realtime crowd-powered interfaces.
\newblock In {\em Proceedings of the 24th Annual ACM Symposium on User
  Interface Software and Technology}, UIST '11, pages 33--42, New York, NY,
  USA, 2011. ACM.

\bibitem{friendsourcing}
Michael~S. Bernstein, Desney Tan, Greg Smith, Mary Czerwinski, and Eric
  Horvitz.
\newblock Personalization via friendsourcing.
\newblock {\em ACM Trans. Comput.-Hum. Interact.}, 17(2):6:1--6:28, May 2008.

\bibitem{crowdsearcher}
Alessandro Bozzon, Marco Brambilla, and Stefano Ceri.
\newblock Answering search queries with crowdsearcher.
\newblock In {\em Proceedings of the 21st International Conference on World
  Wide Web}, WWW '12, pages 1009--1018, New York, NY, USA, 2012. ACM.

\bibitem{CSModel}
Daren~C. Brabham.
\newblock Crowdsourcing as a model for problem solving.
\newblock {\em Convergence: The International Journal of Research into New
  Media Technologies}, 14(1):75--90, 2008.

\bibitem{cssmart}
Georgios Chatzimilioudis, Andreas Konstantinidis, Christos Laoudias, and
  Demetrios Zeinalipour-Yazti.
\newblock Crowdsourcing with smartphones.
\newblock In {\em IEEE Internet Computing}, volume~16, pages 36--44. 2012.

\bibitem{AnalyticsonBD}
Alfredo Cuzzocrea, Il-Yeol Song, and Karen~C Davis.
\newblock Analytics over large-scale multidimensional data: the big data
  revolution!
\newblock In {\em Proceedings of the ACM 14th international workshop on Data
  Warehousing and OLAP}, pages 101--104. ACM, 2011.

\bibitem{mapreduce}
Jeffrey Dean and Sanjay Ghemawat.
\newblock Mapreduce: Simplified data processing on large clusters.
\newblock {\em Commun. ACM}, 51(1):107--113, January 2008.

\bibitem{twitter}
Murat Demirbas, Murat~Ali Bayir, Cuneyt~Gurcan Akcora, Yavuz~Selim Yilmaz, and
  Hakan Ferhatosmanoglu.
\newblock Crowd-sourced sensing and collaboration using twitter.
\newblock In {\em World of Wireless Mobile and Multimedia Networks (WoWMoM),
  2010 IEEE International Symposium on a}, pages 1--9. IEEE, 2010.

\bibitem{cswww}
Anhai Doan, Raghu Ramakrishnan, and Alon~Y. Halevy.
\newblock Crowdsourcing systems on the world-wide web.
\newblock {\em Commun. ACM}, 54:86--96, April 2011.

\bibitem{shepherding}
Steven Dow, Anand Kulkarni, Scott Klemmer, and Bj\"{o}rn Hartmann.
\newblock Shepherding the crowd yields better work.
\newblock In {\em Proceedings of the ACM 2012 Conference on Computer Supported
  Cooperative Work}, CSCW '12, pages 1013--1022, New York, NY, USA, 2012. ACM.

\bibitem{IntersBD}
Danyel Fisher, Rob DeLine, Mary Czerwinski, and Steven Drucker.
\newblock Interactions with big data analytics.
\newblock {\em interactions}, 19(3):50--59, May 2012.

\bibitem{crowddb}
Michael~J. Franklin, Donald Kossmann, Tim Kraska, Sukriti Ramesh, and Reynold
  Xin.
\newblock Crowddb: Answering queries with crowdsourcing.
\newblock In {\em Proceedings of the 2011 ACM SIGMOD International Conference
  on Management of Data}, SIGMOD '11, pages 61--72, New York, NY, USA, 2011.
  ACM.

\bibitem{FIM}
G{\"o}sta Grahne and Jianfei Zhu.
\newblock Fast algorithms for frequent itemset mining using fp-trees.
\newblock {\em Knowledge and Data Engineering, IEEE Transactions on},
  17(10):1347--1362, 2005.

\bibitem{RepSurvey}
Kevin Hoffman, David Zage, and Cristina Nita-Rotaru.
\newblock A survey of attack and defense techniques for reputation systems.
\newblock {\em ACM Comput. Surv.}, 42:1:1--1:31, December 2009.

\bibitem{csfirst}
Jeff. Howe.
\newblock The rise of crowdsourcing.
\newblock {\em Wired}, June 2006.

\bibitem{AleksRep}
Aleksandar Ignjatovic, Norman Foo, and Chung~Tong Lee.
\newblock An analytic approach to reputation ranking of participants in online
  transactions.
\newblock In {\em Proceedings of the 2008 IEEE/WIC/ACM International Conference
  on Web Intelligence and Intelligent Agent Technology - Volume 01}, pages
  587--590, Washington, DC, USA, 2008. IEEE Computer Society.

\bibitem{intrinsic}
Aniket Kittur Boris Smus Jim Laredoc Maja~Vukovic Jakob~Rogstadius,
  Vassilis~Kostakos.
\newblock An assessment of intrinsic and extrinsic motivation on task
  performance in crowdsourcing.
\newblock In {\em Proceeding of the Fifth International AAAI Conference on
  Weblogs and Social Media}. AAAI, 2011.

\bibitem{AMTOpportunities}
Natala J.~Menezes Jenny J.~Chen and Adam~D. Bradley.
\newblock Opportunities for crowdsourcing research on amazon mechanical turk.
\newblock In {\em Proceeding of The CHI 2011 Workshop on Crowdsourcing and
  Human Computation}, May 2011.

\bibitem{crowdweaver}
Aniket Kittur, Susheel Khamkar, Paul Andr{\'e}, and Robert Kraut.
\newblock Crowdweaver: visually managing complex crowd work.
\newblock In {\em Proceedings of the ACM 2012 conference on Computer Supported
  Cooperative Work}, pages 1033--1036. ACM, 2012.

\bibitem{turkomatic}
Anand~P Kulkarni, Matthew Can, and Bjoern Hartmann.
\newblock Turkomatic: automatic recursive task and workflow design for
  mechanical turk.
\newblock In {\em CHI'11 Extended Abstracts on Human Factors in Computing
  Systems}, pages 2053--2058. ACM, 2011.

\bibitem{humanOCR}
Greg Little and Yu-an Sun.
\newblock Human ocr: Insights from a complex human computation process.
\newblock In {\em Workshop on Crowdsourcing and Human Computation, Services,
  Studies and Platforms, ACM CHI}. Citeseer, 2011.

\bibitem{bigdatareport}
James Manyika, Michael Chui, Brad Brown, Jacques Bughin, Richard Dobbs, Charles
  Roxburgh, and Angela~H. Byers.
\newblock {Big data: The next frontier for innovation, competition, and
  productivity}, May 2011.

\bibitem{sortsnjoins}
Adam Marcus, Eugene Wu, David Karger, Samuel Madden, and Robert Miller.
\newblock Human-powered sorts and joins.
\newblock {\em Proc. VLDB Endow.}, 5(1):13--24, September 2011.

\bibitem{mainWWW}
Arjun Mukherjee, Bing Liu, and Natalie Glance.
\newblock Spotting fake reviewer groups in consumer reviews.
\newblock In {\em Proceedings of the 21st international conference on World
  Wide Web}, pages 191--200. ACM, 2012.

\bibitem{elisa2010curation}
Qun Ni and Elisa Bertino.
\newblock Credibility-enhanced curated database: Improving the value of curated
  databases.
\newblock In {\em ICDE}, pages 784--795, 2010.

\bibitem{BigdataLake}
Daniel~E O'Leary.
\newblock Embedding ai and crowdsourcing in the big data lake.
\newblock {\em Intelligent Systems, IEEE}, 29(5):70--73, 2014.

\bibitem{PigLatin}
Christopher Olston, Benjamin Reed, Utkarsh Srivastava, Ravi Kumar, and Andrew
  Tomkins.
\newblock Pig latin: A not-so-foreign language for data processing.
\newblock In {\em Proceedings of the 2008 ACM SIGMOD International Conference
  on Management of Data}, SIGMOD '08, pages 1099--1110, New York, NY, USA,
  2008. ACM.

\bibitem{crowdscreen}
Aditya~G. Parameswaran, Hector Garcia-Molina, Hyunjung Park, Neoklis Polyzotis,
  Aditya Ramesh, and Jennifer Widom.
\newblock Crowdscreen: Algorithms for filtering data with humans.
\newblock In {\em Proceedings of the 2012 ACM SIGMOD International Conference
  on Management of Data}, SIGMOD '12, pages 361--372, New York, NY, USA, 2012.
  ACM.

\bibitem{Sawzall}
Rob Pike, Sean Dorward, Robert Griesemer, and Sean Quinlan.
\newblock Interpreting the data: Parallel analysis with sawzall.
\newblock {\em Scientific Programming}, 13(4):277--298, 2005.

\bibitem{growingfield}
Alexander~J. Quinn and Benjamin~B. Bederson.
\newblock Human computation: a survey and taxonomy of a growing field.
\newblock In {\em Proceedings of the 2011 annual conference on Human factors in
  computing systems}, CHI '11, pages 1403--1412, New York, NY, USA, 2011. ACM.

\bibitem{cloudTax}
Bhaskar~Prasad Rimal, Eunmi Choi, and Ian Lumb.
\newblock A taxonomy and survey of cloud computing systems.
\newblock In {\em INC, IMS and IDC, 2009. NCM'09. Fifth International Joint
  Conference on}, pages 44--51. Ieee, 2009.

\bibitem{ibmBigData}
Philip Russom et~al.
\newblock Big data analytics.
\newblock {\em TDWI Best Practices Report, Fourth Quarter}, 2011.

\bibitem{sherifbigdata2011}
Sherif Sakr, Anna Liu, Daniel~M Batista, and Mohammad Alomari.
\newblock A survey of large scale data management approaches in cloud
  environments.
\newblock {\em Communications Surveys \& Tutorials, IEEE}, 13(3):311--336,
  2011.

\bibitem{bda2012}
S~Srinivasa and V~Bhatnagar.
\newblock Big data analytics.
\newblock In {\em Proceedings of the First International Conference on Big Data
  Analytics BDA}, pages 24--26, 2012.

\bibitem{wallmart}
Chong Sun, Narasimhan Rampalli, Frank Yang, and AnHai Doan.
\newblock Chimera: Large-scale classification using machine learning, rules,
  and crowdsourcing.
\newblock {\em Proceedings of the VLDB Endowment}, 7(13), 2014.

\bibitem{IEEE2102Survey}
Y.(. Sun and Y.~Liu.
\newblock Security of online reputation systems: The evolution of attacks and
  defenses.
\newblock {\em Signal Processing Magazine, IEEE}, 29(2):87 --97, march 2012.

\bibitem{crowdwisdombook}
James Surowiecki.
\newblock {\em The Wisdom of Crowds}.
\newblock Anchor Books, 2005.

\bibitem{CSforEnter}
M.~Vukovic.
\newblock Crowdsourcing for enterprises.
\newblock In {\em Services - I, 2009 World Conference on}, pages 686 --692,
  july 2009.

\bibitem{towardCS}
Maja Vukovic and Claudio Bartolini.
\newblock Towards a research agenda for enterprise crowdsourcing.
\newblock In Tiziana Margaria and Bernhard Steffen, editors, {\em Leveraging
  Applications of Formal Methods, Verification, and Validation}, volume 6415 of
  {\em Lecture Notes in Computer Science}, pages 425--434. Springer Berlin /
  Heidelberg, 2010.

\bibitem{peoplecloude}
Maja Vukovic, Mariana Lopez, and Jim Laredo.
\newblock Peoplecloud for the globally integrated enterprise.
\newblock In {\em Service-Oriented Computing. ICSOC/ServiceWave 2009
  Workshops}, volume 6275 of {\em Lecture Notes in Computer Science}, pages
  109--114. Springer Berlin / Heidelberg, 2010.

\bibitem{turf}
Gang Wang, Christo Wilson, Xiaohan Zhao, Yibo Zhu, Manish Mohanlal, Haitao
  Zheng, and Ben~Y Zhao.
\newblock Serf and turf: crowdturfing for fun and profit.
\newblock In {\em Proceedings of the 21st international conference on World
  Wide Web}, pages 679--688. ACM, 2012.

\bibitem{crowdER}
Jiannan Wang, Tim Kraska, Michael~J. Franklin, and Jianhua Feng.
\newblock Crowder: Crowdsourcing entity resolution.
\newblock {\em Proc. VLDB Endow.}, 5(11):1483--1494, July 2012.

\bibitem{reptrap}
Yafei Yang, Qinyuan Feng, Yan~Lindsay Sun, and Yafei Dai.
\newblock Reptrap: a novel attack on feedback-based reputation systems.
\newblock In {\em Proceedings of the 4th international conference on Security
  and privacy in communication netowrks}, page~8. ACM, 2008.

\bibitem{zaharia2008improving}
Matei Zaharia, Andy Konwinski, Anthony~D Joseph, Randy~H Katz, and Ion Stoica.
\newblock Improving mapreduce performance in heterogeneous environments.
\newblock In {\em OSDI}, volume~8, page~7, 2008.

\end{thebibliography}

\end{document}